\documentclass[aps, amssymb, amsmath, superscriptaddress, prb, 
twocolumn] {revtex4-2}

\usepackage{graphicx}
\usepackage{color}
\usepackage{amsmath}
\usepackage{enumitem}
\usepackage{amssymb}
\usepackage{hyperref}
\usepackage{cancel}
\usepackage{ulem}
\usepackage{multirow}

\usepackage{pifont}

\newcommand{\be}{\begin{equation}}
	\newcommand{\ee}{\end{equation}}
\newcommand{\bea}{\begin{eqnarray}}
	\newcommand{\eea}{\end{eqnarray}}

\newcommand{\tr}{{\rm \, tr\,}}

\renewcommand{\Im}{{\rm \, Im\,}}

\renewcommand{\vec}[1]{{\boldsymbol #1}}

\renewcommand\vec[1]{\ensuremath\boldsymbol{#1}} 

\usepackage{amsfonts, relsize, color}
\usepackage{graphicx}
\usepackage{color}
\usepackage{comment}

\usepackage{xcolor}
\hypersetup{
	colorlinks,
	linkcolor={red!50!black},
	citecolor={green!50!black},
	urlcolor={blue!50!black}
}
\usepackage{booktabs} 
\begin{document}
	
\title{
Particle-Hole Ghost Interference in Superconductors}

\author{Archisman Panigrahi}\affiliation{Department of Physics, Massachusetts Institute of Technology, 77 Massachusetts Ave, Cambridge 02139, USA}
\author{Vladislav Poliakov}\affiliation{Department of Physics, Massachusetts Institute of Technology, 77 Massachusetts Ave, Cambridge 02139, USA}
\author{Leonid Levitov}\affiliation{Department of Physics, Massachusetts Institute of Technology, 77 Massachusetts Ave, Cambridge 02139, USA}

\date{\today}

\begin{abstract}
Mirror-assisted optical interference can improve the fidelity of Young’s double-slit interference. Here we discuss an electron analogue: particle-hole interference in superconductors produced by a single impurity near a line defect, terrace edge, or phase boundary. Quasiparticle waves scattered directly from the impurity interfere with waves reflected by the boundary, generating a “ghost” interference pattern that combines conventional $2k_F$ Friedel oscillations with additional hyperbolic fringes. Compared to the recently studied two-impurity Young’s interference, this effect appears already at first order in the impurity potential and is therefore parametrically stronger. The resulting spatial modulation extends beyond $\lambda_F/2$ and is directly sensitive to the quasiparticle structure of the paired state, including possible Fermi-surface anisotropy of the superconducting order parameter. These findings point to boundary-assisted impurity interference as a robust local probe of superconducting electronic order, with clear signatures accessible to STM/STS measurements.
\end{abstract}

\maketitle


Using local probes to investigate exotic superconductivity has been an active research direction in recent years~\cite{Balatsky2006,fischer2007stm,zeldov2013squid,yazdani2021matbg,
yin2021stm,yazdani2024moire}. These probes have been used successfully to probe Friedel oscillations, which arise from $2k_F$ backscattering of quasiparticles by localized impurities and encode detailed information about Fermi-surface geometry, topology, and Berry phase, making them a probe of choice for the electronic structure of metals~\cite{Bena2016}. 
In superconductors, however, the relation between Friedel oscillations and the paired state remains less well understood. 
Nevertheless, impurity-induced modulations and quasiparticle interference have been widely used to probe superconducting order, with notable successes in multiband systems such as the iron pnictides, where they have been used to constrain the relative sign of the gap on different Fermi-surface sheets~\cite{Hirschfeld2015,Zhang2014}. 
In conventional single-band superconductors, impurity-induced oscillations of both the order parameter and the local density of states are also well established, although the prevailing view has been that superconductivity primarily suppresses normal-state Friedel oscillations beyond the coherence-length scale~\cite{Stosiek2022}.

\begin{figure}[t]
    \centering
    \includegraphics[width=0.99\linewidth]{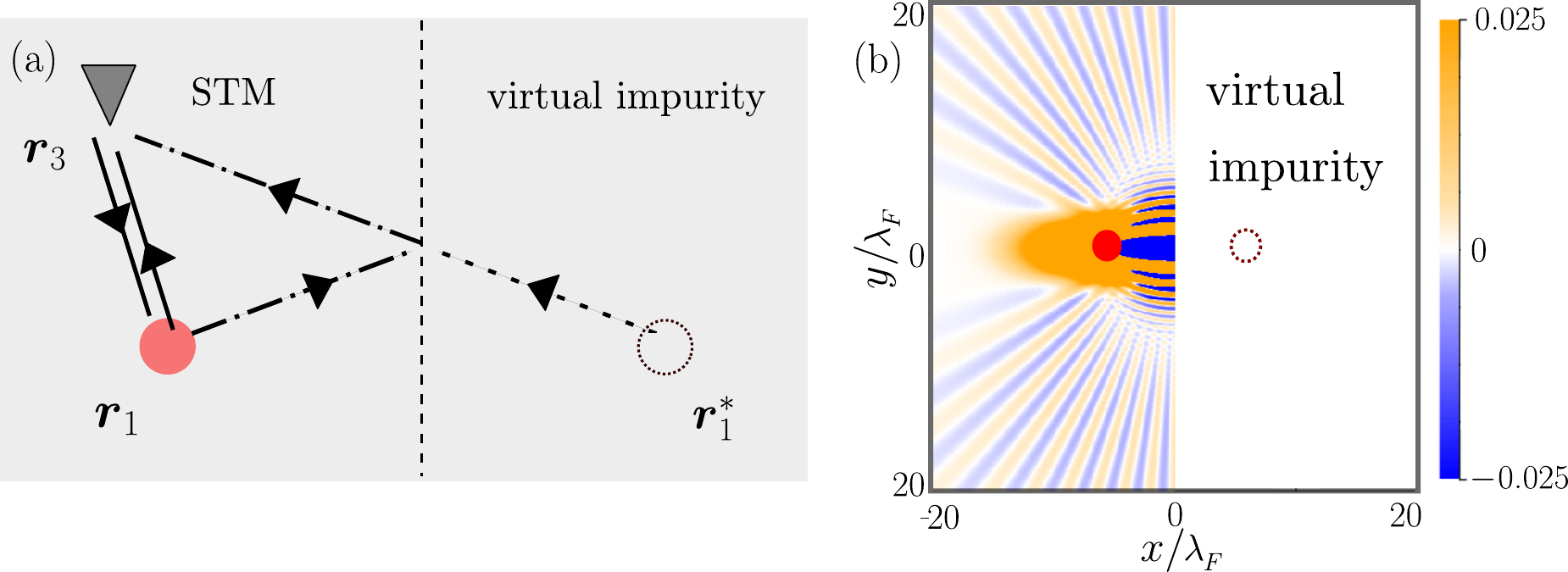}
    \caption{(a) 
    Boundary-assisted interference effect in the TDOS of an impurity and its mirror image due to scattering at a domain wall. Here, $\vec r_1$ is the position of impurity, $\vec r_1^*$ is the position of `virtual impurity' (`ghost impurity'), and the STM tip is shown at position $\vec r_3$. (b) Fourier filtered TDOS (Eq.\eqref{eq:TDOS-zero-temp}) for a $d+id$ superconductor at $eV = 1.01 \Delta$ displays hyperbolic Young's interference patterns. The $x$ and the $y$ coordinates specify the position of the moving STM tip. The orange and blue colors represent positive and negative sign of TDOS, which is plotted in the units of $\frac{4U k_F e^2 |t_{\rm STM}|^2 \nu_0}{\pi v_F^2 \lambda_F}$. Here we considered reflection coefficient $\lambda = 0.2$, and coherence length $\xi = \hbar v_F/\Delta = 1000 \lambda_F$.}
    \label{fig:mirror-image}
\end{figure}

The present work stems from the attempts to identify geometry in which Friedel oscillations are directly sensitive to angle-dependent superconducting order parameter at the Fermi surface\cite{Ding2023,Panigrahi2025}. To that end, we consider a geometry, in which an impurity is located near a line defect, terrace edge, or phase boundary. In this setting, the impurity's mirror image can act as a virtual scattering center, so the quasiparticle waves reflected by the line defect interfere with the direct wave reflected  from the impurity. This geometry can be viewed as an electronic analogue of Lloyd's mirror geometry, first described in optics in 1834~\cite{Lloyd1834}. In the optical problem, Lloyd geometry enhances the visibility of Young's interference fringes; here, we show that an analogous enhancement occurs for particle-hole interference in superconductors. The resulting pattern combines ordinary $2k_F$ Friedel oscillations with additional mirror-generated fringes, which have the geometry similar to the hyperbolic particle-hole Young's interference from two impurities, however, occurring due to a single impurity positioned near a line defect or phase boundary. 

Importantly, as we will see, this `ghost interference' effect arises at first order of perturbation theory in the impurity potential and, therefore, it is considerably stronger than the two-impurity Young's interference studied in Refs.\cite{Ding2023,Panigrahi2025}, which is a second-order effect. Furthermore, the resulting hyperbolic fringes occur at the lengthscales greater than the Friedel oscillations periodicity, in complete analogy with the predictions of Refs.\cite{Ding2023,Panigrahi2025}. This property, as well as the unusual strength of this interference effect will facilitate its detection stproducing a spatial interference structure that can be naturally accessed 
by STM/STS probes.

This Lloyd-mirror picture is also closely connected to the classic theory of Friedel oscillations and to the STM literature on boundary-induced standing waves at step edges and terraces, including Cu(111) surface-state imaging and confined Friedel oscillations on Au(111) terraces~\cite{Friedel1958,Hormandinger1994,Sotthewes2021}. In this boundary-assisted geometry, the defect--boundary pair acts as a coherent interferometer whose fringe pattern encodes the underlying quasiparticle structure of the metal or superconductor.

While two-impurity interference in STM signal can be utilized to detect phase winding of a topological superconductor, it is difficult to realize experimentally. Here, we propose as alternative mechanism, where the boundary creates the effect of a `ghost' impurity in the Green's function. Due to the effect of the boundary, the quasiparticle is reflected back from it (see Fig.\ref{fig:mirror-image}), and the Green's function is modified as,
\begin{equation}\label{eq:G-boundary}
    \tilde G^{R(0)}_{\rm edge}(\omega, \vec r, \vec r') =  G^{R(0)}_{\vec r-\vec r{'}} (\omega)-  G^{R(0)}_{\vec r-\vec {r'}^*}(\omega)
\end{equation}
where $\vec {r'}^*$ denotes the mirror image of the point $\vec {r'}$. The $-$ve sign multiplying the mirror Green's function arises from the boundary conditions. In case of a domain wall (rather than a metal-vacuum boundary) or a line defect, we can phenomenologically write,
\begin{equation}\label{eq:G-boundary-impurity}
    \check{G}^{(0)}_{\rm domain}(\omega, \vec r, \vec r') =  G^{R(0)}_{\vec r-\vec r{'}} (\omega)- \lambda  G^{R(0)}_{\vec r-\vec {r'}^*}(\omega)
\end{equation}
where $\lambda$ is a dimensionless parameter describing the amplitude of the wave reflected at the domain wall.

Then, for a quasiparticle scattered by a single impurity at $\vec r_1$, there is an additional contribution due to reflection at the boundary, which effectively mimics scattering by a mirror image impurity at $\vec r_1^*$ (see Fig.\ref{fig:mirror-image}), and the full impurity-induced Green's function becomes,
\begin{widetext}
\begin{equation}\label{eq:domain-wall-imp-G}
\begin{aligned}
    \check G^R(\omega, \vec r, \vec r') &=  \check G^{R(0)}(\omega, \vec r, \vec r') + \check G^{R(0)}(\omega, \vec r, \vec r_1) (U \tau_z) \check G^{R(0)}(\omega, \vec r_1, \vec r')\\
    &= \check G^{R(0)}(\omega, \vec r, \vec r') + G^{R(0)}_{\vec r-\vec r_1} (\omega) (U \tau_z) G^{R(0)}_{\vec r_1-\vec r{'}} (\omega) -\lambda \left[{G^{R(0)}_{\vec r- \vec r_1}}(\omega) (U \tau_z) {G^{R(0)}_{\vec r_1^*- \vec r'}} +  {G^{R(0)}_{\vec r- \vec r_1^*}} (U \tau_z) {G^{R(0)}_{\vec r_1- \vec r'}}\right] + \cdots
\end{aligned}
\end{equation}
\end{widetext}
where `$\cdots$' denote higher-order contributions. The first term is the original Green's function near the domain, the second term (order $\mathcal{O}(U)$) is the impurity modified bulk Green's function describing regular Friedel oscillations, and the last term (order $\mathcal{O}(\lambda U)$) describes the quasiparticle propagation due to the combined effect of the impurity and the domain boundary. It is the last term which gives rise to the Lloyd-mirror effect, which is the central result of the paper.

\section{Spatial dependence of the TDOS near and impurity and line defect}

To compute the effect of an impurity placed near a domain boundary, on the tunneling density of states (TDOS) in a superconductor, we will first compute the impurity-modified Green's function, and then relate it to the experimentally measured TDOS.
The BdG Hamiltonian with a gap function $\Delta(\vec{k})$ takes the form,
\begin{equation}
    H^0(\vec k) = \begin{pmatrix}
    \xi_{\vec k} & \Delta(\vec k) (i\sigma_y)\\
    \Delta^*(\vec k) (-i \sigma_y) & -\xi_{-\vec k}
\end{pmatrix}
\end{equation}
where $\xi_{\vec k} =  v_F (k-k_F)$ is quasiparticle dispersion in the normal metal.

The real-space Green's function is given by
\begin{equation}
    G^{R(0)}_{\vec r - \vec r'}(\omega) = \int \frac{d^2 \vec k}{(2\pi)^2} \frac{e^{i \vec k \cdot (\vec r - \vec r')}}{(\omega + i \eta) - H^0(\vec k)}.
\end{equation}


From this Green's function describing the propagation of quasiparticles away from an impurity or a domain boundary, the impurity-modified Green's function $\check G^R(\omega, \vec r, \vec r')$ near the domain boundary can be computed, as described in Eq. \eqref{eq:domain-wall-imp-G}.

The electronic spectral function $A_{e}(\vec r_3, V)$ can be computed from the impurity-modified Green's function,
\begin{equation}
    A_{e}(\vec r_3, V) = -\frac{1}{\pi}\tr  \Im \left[\left(\frac{1+\tau_z}{2}\right) \check G^R(eV, \vec r, \vec r)\right],
\end{equation}
which is related to the tunneling density of states $\frac{dI}{dV}$ (TDOS) as,
\begin{equation}\label{eq:working-formula-dIdV}
 \frac{dI (\vec r_3)}{dV} 
 = -4 \pi e^2 |t_{\rm STM}|^2 \nu_0 
\int d \omega A_e(\vec r_3, \omega) f'(\omega - eV),
\end{equation}
where `$\tr$' involves tracing over the spin and particle-hole degrees of freedom, $\nu_0$ is the density of states at the Fermi level, $t_{\rm STM}$ is the tunneling amplitude between the STM tip and the superconductor, and $f(x) = \frac{1}{e^{\beta x} + 1}$ is the Fermi function. 
At low temperature, the derivative of the Fermi function is a Dirac delta, producing
\begin{equation}\label{eq:TDOS-zero-temp}
   \frac{dI (\vec r_3)}{dV} = 4 \pi e^2 |t_{\rm STM}|^2 \nu_0  A_{e}(\vec r_3, eV).
\end{equation}

For a bias voltage $V$ below the gap $\Delta$, the trace of the Green's function is purely real, and the density of states is zero. We discuss the situation when the bias voltage is above the gap. To prevent confusion regarding different branch cuts, we only present results for a positive bias voltage. For negative bias, the calculation can be done in a similar manner.

When the distance between the tip to impurity is much larger than the Fermi wavelength (Eilenberger limit), the Green's function assumes the form,

\begin{align}\label{eq:general-SC-green's-function}
	G^{R(0)}(\omega,\vec r) &= e^{i\gamma_r} g_+(\vec r) + e^{-i\gamma_r} g_-(\vec r)\\ \nonumber
	g_+(\vec r) = A_{\vec r} \tau_0 &+ B_{\vec r} \tau_z + {C_1}_{\vec r} \tau_{+} (i\sigma_y) + {C_2}^\dagger_{\vec r} \tau_{-} (-i\sigma_y)\\ \nonumber
	g_-(\vec r) = A_{-\vec r} \tau_0 &- B_{-\vec r} \tau_z + {C_1}_{-\vec r} \tau_{+} (i\sigma_y) + {C_2}^\dagger_{-\vec r} \tau_{-} (-i\sigma_y)
\end{align}
where we denote 
\[
(A_{\vec r},B_{\vec r}, {C_1}_{\vec r},{C_2}_{\vec r}) = D_{\vec r} \left(\frac{i\omega}{\Omega_{\vec r}}, i, \frac{i\Delta(\vec k_{\vec r})}{\Omega_{\vec r}},\frac{i\Delta^*(\vec k_{\vec r})}{\Omega_{\vec r}}\right),
\] 
\[ \Omega_{\vec r} = \sqrt{(\omega + i\eta)^2 - |\Delta(\vec k_{\vec r})|^2}, 
\quad \tau_{\pm} = \frac{\tau_x \pm i \tau_y}{2},\] 
\[ D_{\vec r} = -\sqrt{\frac{k_F}{8 \pi v_F^2 r}} {e^{i r\frac{\Omega_{\vec r}}{v_F}}}.
\] 
The quantities $g_{+}(\vec r)$ and  $g_{-}(\vec r)$ describe the electron-like and hole-like propagators, respectively.
Here $\vec r$ denotes $(\vec r-\vec r')$, and $\gamma_r = (k_F r - \frac{\pi}{4})$ is distance in Fermi wavelength units; $\Delta(\vec k_{\vec r})$ denotes $\Delta(\vec k)$ with $\vec k$ collinear with $\vec r$ 
(and $|\vec k|=k_F$). $\eta$ is an infinitesimally small positive real number. For a spin-polarized triplet superconductors, we have to replace the spin matrices $\pm i \sigma_y$ with unity.
As an illustration, we consider a spin-singlet $d+id$ superconductor, with
\begin{equation}
    \Delta(\vec k) = \Delta e^{2 i \theta_{\vec k}} = \Delta \left(\frac{k_x + i k_y}{k_F}\right)^2.
\end{equation}

We find that for bias voltage below the gap, there is no Friedel oscillation at zero temperature, as $A_e(\vec r_3,\omega) = 0$ for $|\omega|< |\Delta|$.
For a bias voltage above the gap ($eV>\Delta$), first order contribution from an impurity of strength $U$ at $\vec r_1$ takes the form,
\begin{equation}\label{eq:Friedel}
    \begin{aligned}
        A_e(\vec r_3, eV) =A^{(1)}  \frac{eV  \cos{\left(\frac{2 \sqrt{(eV)^2 - |\Delta|^2}r_{31}}{ v_F}\right)}}{\sqrt{(eV)^2 - |\Delta|^2)}} 
        \frac{\cos(2 k_F r_{31}) }{r_{31}} ,
    \end{aligned}
\end{equation}
with $A^{(1)} = \frac{-U}{\pi} (\frac{k_F}{\pi v_F^2})$.
The only difference between the Friedel oscillation in a superconductor and that in normal metal is that, there is no STM signal below the gap in a superconductor, and just above the gap, there is a divergent density of states, as expected from BCS theory. Moreover, the bias-voltage dependent oscillations behave as $\cos{\left(\frac{2 \sqrt{(eV)^2 - |\Delta|^2}r_{31}}{\hbar v_F}\right)}$ compared to $\cos{\left(2 e V r_{31}/\hbar v_F\right)}$ dependence in a normal metal. Notably, bias-voltage dependent Tomasch oscillations \cite{Tomasch1965, Tomasch1966} arise at an energy $\omega_0 = \frac{2\pi v_F}{2 r_{31}} \sim \Delta \frac{\xi}{2 r_{31}}$, where $\xi \gg r_{31}$ is the coherence length of the superconductor.
At such large voltages, the effect of the superconducting gap is washed out and the bias-voltage dependence is not distinguishable from its normal metal counterpart.

The ghost interference contribution to the Green's function for a $d+id$ superconductor, due to the combined effect of an impurity and the domain wall takes the form,
\begin{widetext}
\begin{equation}\label{eq:ghost-interference-contribution-hyperbola}
A_e(\vec r_3, eV)|_{\rm ghost} = \frac{\lambda U \Delta^2}{\omega^2-\Delta^2} {\frac{ k_F}{\pi^2 v_F^2 \sqrt{r_{31}r_{31^*}}}}\cos(k_F (r_{31}-r_{31^*}))\sin^2(\theta_{31} - \theta_{31^*}) \sin{\left(\frac{\sqrt{(eV)^2-\Delta^2}(r_{31}+r_{31^*})}{v_F}\right)} + \cdots
\end{equation}
\end{widetext}
where the `$\cdots$' contain the particle-particle contributions, which behave like Friedel oscillations, and $\theta_{ij}$ denotes the angle between $\vec r_i - \vec r_j$ and the $x$ axis.
Clearly, the effect is strongest when $eV$ is just above the gap $\Delta$.
As discussed previously in the context of regular Friedel oscillations, the bias-voltage dependent oscillatory exponential will generate an analog of Tomasch oscillations, but at frequencies much larger than $\Delta$ where the quasiparticles are agnostic of the gap function.

\section{Fourier filtering of 
quasiparticle interference patterns}

The zeroth order contribution is the unperturbed Green's function $G^{R(0)}_{\vec r-\vec r'}$, which produces a uniform background. Since the components of the unperturbed Green's function behave as $e^{\pm i k_F |\vec r - \vec r'|}$, the first order contribution at a large distance $R$ from an impurity behaves as $\frac{\cos (2 k_F R)}{R}$, which produces a peak at $2 k_F$ when Fourier transformed to momentum space.

 The contribution to TDOS due to the ghost interference features factors such as $\sin(k_F (r_{31}-r_{31^*}))$, which depend on the path-difference between the tip-to-impurity and tip-to-ghost-impurity trajectories, and produces the hyperbolic Young's interference patterns mentioned earlier.
Such features arise due to the coherent inter-conversion of particles (which propagate as $e^{i k_F r}$) and holes (which propagate as $e^{-i k_F r}$) in a superconductor \cite{Ding2023, Panigrahi2025}, as the Bogoliubov quasiparticles are made of linear combinations of both particles and holes. In addition, the second order contribution to the Green's function also features particle-particle contributions of the form  $\sin(k_F (r_{23}+r_{31}))$, which are also present in a normal metal and occur due to particle-particle contributions, and these form an ellipse-like patterns in real space. Such ellipse-like patterns asymptotically become concentric circles (with wavenumber $2 k_F$) at large distances, as the two impurities effective
behave as a single one.

In Fig.\ref{fig:mirror-image}(b), the Fourier-filtered ghost interference pattern is displayed after masking the large momentum modes at $2 k_F$ which contribute to Friedel oscillations. We first numerically compute the full spectral function using Eq.\eqref{eq:domain-wall-imp-G} (not just the particle-hole contribution in Eq.\eqref{eq:ghost-interference-contribution-hyperbola}), and then Fourier filter it to isolate the particle-hole contribution in ghost interference.

\section{Discussions and Conclusion}

We showed that an impurity near a line defect, terrace edge or domain boundary realizes an electronic analogue of Lloyd-mirror geometry, where quasiparticles scattered directly from an impurity interfere with those reflected by the boundary. The boundary modifies the Green's function in a way that the propagation of quasiparticles is effectively controlled by the original impurity, and its mirror image, which we dub the `ghost' impurity (see Fig.\ref{fig:mirror-image}(a)). As a result, a single physical scatterer can produce a two-source interference pattern. Crucially, this contribution is first order in the impurity potential $U$ (more precisely, at order $\lambda U$ where $\lambda$ is the reflection coefficient at the domain wall). The analogous hyperbolic Young's interference fringes studied in Refs.\cite{Ding2023, Panigrahi2025} arise in second order $\mathcal{O}(U^2)$ which is perturbatively smaller than the ghost interference. Being parametrically stronger, the ghost interference should be easier to resolve in STM experiments.

The resulting spatial pattern of TDOS contains two physically distinct contribution, which are separable in both real and momentum space. On its own, the impurity produces Friedel oscillations (Eq.\eqref{eq:Friedel}) that produce concentric rings of wavenumber $2 k_F$ around the impurity. The combined effect of the impurity and the domain wall contribute a term that depends on the the sum and difference of the path lengths of the tip-to-impurity and tip-to-image trajectories. The former produces hyperbolic Young's interference patterns, while the latter produces ellipses whose foci are the original impurity and its image counterpart. Due to their rapid variation, both the Friedel oscillation and the ellipse-like contribution occur at a large wavenumber $2 k_F$, whereas the Young's interference fringes correspond to a much smaller wavenumber in the Fourier transform of the spatial TDOS pattern, allowing them to be separated with Fourier filtering.

The reflecting boundary is treated phenomenologically with a constant reflection amplitude $\lambda$ in Eq.\eqref{eq:G-boundary}. In a realistic system, $\lambda$ and its energy dependence will depend on the nature of the reflecting boundary.
This raises the possibility of using boundary-assisted interference to characterize domain boundaries of different superconducting order parameters, including the role of topological edge modes in particle-hole interference.

The boundary-assisted interference effect is appealing from several points of view. On the theoretical side, it arises at first order in the impurity potential and is therefore expected to be stronger than the two-impurity interference studied in Refs.~\cite{Ding2023, Panigrahi2025}. On the experimental side, it does not require controlled placement of two impurities, unlike the two-impurity setup of Refs.~\cite{Ding2023, Panigrahi2025}. Furthermore, it can be used to probe quasiparticle dynamics at phase boundaries, such as interfaces between regions with different chirality or topology in strongly correlated systems, assisted by one-dimensional collective edge modes localized at and propagating along the system’s edges.

%


This work was performed in part at the Aspen Center for Physics, which is supported by National Science Foundation grant PHY-2210452. It was also supported in part by the U.S.-Israel Binational Science Foundation (BSF).



\end{document}